\documentclass[preprint,12pt]{elsarticle}
\usepackage{amssymb}
\newcounter{bla}

  \journal{Computer Physics Communications}
  \usepackage[version=4]{mhchem}
  \usepackage{physics}
  \usepackage[table]{xcolor}
  \usepackage{amsmath}
  \usepackage{amssymb}
  \usepackage{mathtools}
  \usepackage{amsbsy}
  \usepackage[scr=boondox]{mathalpha}
  \usepackage{mathdots}
  \usepackage{siunitx} 
  \usepackage{nicefrac}
  \usepackage{listings}
  \usepackage{hyperref}
  \usepackage{empheq}
  \usepackage{comment}
  \usepackage{bm}

  \lstdefinestyle{simplestyle}{  
  commentstyle=\color{gray}\itshape,
  keywordstyle=\color{blue},
  basicstyle=\ttfamily\footnotesize,
  breakatwhitespace=false,         
  breaklines=true,                 
  captionpos=b,                    
  keepspaces=true,                                   
  numbersep=5pt,                  
  showspaces=false,                
  showstringspaces=false,
  showtabs=false,                  
  tabsize=2
  }

  \newcommand{\peng}{\textsc{Peng}}
  \newcommand{\kB}{k_{\mathrm{B}}}
  \usepackage{accents}
  
  \newcommand{\nodiv}[1]{\ensuremath{\accentset{0}{#1}}}
  \newcommand{\D}{\mathscr{D}}
  \newcommand{\C}{\vect{\mathscr{C}}}
  \newcommand{\vect}[1]{\bm{#1}}
  \newcommand{\mat}[1]{\bm{\mathsf{#1}}}
  \newcommand{\T}{\mathsf{T}}
  \renewcommand{\tilde}{\widetilde}
  
\begin{document}

  \begin{frontmatter}



    \title{\peng{}: \\A program for transport properties of low-density binary gas mixtures}


    \author[a]{Yu Zhai}
    \author[a]{You Li}
    \author[a,b]{Hui Li\corref{author}}
    \author[b]{Frederick R. W. McCourt}

    \cortext[author] {Corresponding author.\\\textit{E-mail address:} prof\_huili@jlu.edu.cn}
    \address[a]{Institute of Theoretical Chemistry, College of Chemistry, Jilin University, 2519 Jiefang Road, Changchun, 130023, China}
    \address[b]{Department of Chemistry, University of Waterloo, Waterloo, ON, N2L 3G1 Canada}

    \begin{abstract}
      The fundamental properties of molecules bridge experiment and theory.
      Transport properties (diffusion, thermal diffusion, thermal conductivity and viscosity) 
      of binary mixtures are measurable in experiments, 
      and well-defined in theory, but difficult to compute with high accuracy. 
      In addition to high-accuracy inter-molecular potential energy curves (PECs), 
      a reliable and high-order solution program that compute the properties 
      based on the PECs is required.
      In this work, we present a computer program called \peng{} that performs the collision integration numerically,
      and solves the Boltzmann equation in Chapman--Enskog fashion.
      The program has been devised to perform both parts of the solution procedure to arbitrary order,
      so that no hard-coded limitation will prevent a user from computing at higher precision, 
      except the amount of RAM and the required computational time.
      \peng{} is well-designed in an Object-Oriented Programming (OOP) fashion, 
      which make the program clear and easy to modify.
      In addition to the end-user oriented program,
      \peng{} is also compiled as a dynamic shared library
      that may readily be extended and embedded in users' programs.
    \end{abstract}


  \end{frontmatter}

  % \linenumbers

  % All CPiP articles must contain the following
  % PROGRAM SUMMARY.

  {\bf PROGRAM SUMMARY}
  %Delete as appropriate.

  \begin{small}
    \noindent
    {\em Program Title:} \peng{}

    \noindent
    {\em CPC Library link to program files:} (to be added by Technical Editor)

    \noindent
    {\em Developer's repository link:} (if available) \url{https://github.com/zhaiyusci/peng}

    \noindent
    {\em Code Ocean capsule:} (to be added by Technical Editor)

    \noindent
    {\em Licensing provisions(please choose one):} LGPL 

    \noindent
    {\em Programming language:} C++

    \noindent
    {\em Supplementary material:} 

    \noindent{\em Nature of problem (approx. 50-250 words):}  
    Nowadays, quantum chemistry provides high-accuracy intermolecular 
    interactions potential energy curves (PECs),
    and more and more accurate thermophysical properties of dilute gases 
    can be measured experimentally.  
    It is meaningful to build a bridge between thermophysical properties 
    and PECs, so that people can refine the PECs and learn more about the nature
    of dynamics of gases. 
    For a dilute binary gas mixture, 
    the theory, the Boltzmann equation, 
    and its solution has been available for a long time,
    but fewer numerical solution packages are available for the public.
    An easy-to-use and easy-to-extend software package is required. 
    %Describe the nature of the problem here. \\

    \noindent
    {\em Solution method (approx. 50-250 words):}
    To complete the work, 
    we have created a program that computes the collision integrals ($\Omega^{(\ell,s)}$) 
    for a set of temperatures, following which 
    the thermophysical properties are computed using the Chapman--Enskog solutions
    of the Boltzmann equation. 
    Both parts of the program are written in C++.
    The program is easy to use,
    and a well-designed framework is provided thanks to the Object-oriented design.
    Users with programming experience can readily extend the present work.
    %Describe the method solution here.

    \noindent
    {\em Additional comments including restrictions and unusual features (approx. 50-250 words):}
    In this version, the program is limited 
    to the computation of the properties of mixtures of structureless atoms.
    For polyatomic molecules, additional work need to be done.
    %Provide any additional comments here.

    % \begin{thebibliography}{0}
    %   \bibitem{1}Reference 1         % This list should only contain those items referenced in the                 
    %   \bibitem{2}Reference 2         % Program Summary section.   
    %   \bibitem{3}Reference 3         % Type references in text as [1], [2], etc.
    %     % This list is different from the bibliography at the end of 
    %     % the Long Write-Up.
    % \end{thebibliography}
    % * Items marked with an asterisk are only required for new versions
    % of programs previously published in the CPC Program Library.\\
  \end{small}

  \section{Introduction}
  \label{sec:introduction}

  Transport properties 
  (thermal conductivity, thermal diffusion, diffusion, and viscosity)
  are important thermophysical features for gas mixtures.\cite{Hirschfelder1954}
  As other dynamical phenomena, 
  these properties are directly related to the interactions, 
  \textit{i.e.}, the intermolecular potential energy functions/surfaces/curves (PEFs/PESs/PECs).

  It was unrealistic 
  to compute the high-accuracy thermophysical properties 
  fully from the first principle.
  Computational quantum chemistry nowadays 
  can provide very accurate potential energy within a (relatively) short time.
  Larger memory, higher CPU speed and parallel programming 
  have made the computation impossible in the old days possible.

  Beside the \textit{ab initio} researching paradigm, 
  which is popular among computational scientists, 
  we should also note that no matter how accurate theory is, 
  it is meaningful to refine (or obtain) the theoretical results
  with help from experimental scientists. 
  There is a long history of getting the parameters 
  directly from experiments 
  by fitting the model to measurement.
  \textit{E.g.}, directly fitting PECs to spectra and virial coefficients 
  is possible even for modern potential energy models,
  which have multiple parameters 
  to represent realistic PECs flexibly and accurately.\cite{dPotFit2017}
  Le~Roy and coworkers proposed in Ref.~\cite{Myatt2018} that it is possible 
  to add gas transport properties into the data set to which the model is fitted,
  and they tuned their PECs manually 
  to obtain a globally better description of all four kinds of properties. 

  Apparently, one of necessary steps of all such proposals is to build a
  PEC to transport properties software.
  There has been some\cite{Maitland1981, Barker1964, Taylor1979, OHara1971}, but most of them suffer from one or more 
  problems listed below:
  (1) The programs were composed in an old-fashioned style,
  \textit{e.g.}, out-of-date language/syntax, 
  ``all in main function'' designing which hampered them being embedding in other applications,
  or ``GOTO'' dominated and thus unmaintainable logic.
  Meanwhile, due to an abuse of global variables, 
  it is so difficult to adopt these codes that, in our experience, 
  we must have 
  three copies of the same code for computing $\Omega_{\alpha\beta}^{(\ell, s)}$ 
  (here $\alpha$ and $\beta$ for different species) 
  and compile them with different potential functions ($V_{\alpha\beta}$)
  into dynamically loadable libraries 
  and load them in the following program for binary mixture transport properties.
  (2) Limited precision was hard coded in the programs, 
  \textit{e.g.}, small numbers of quadrature points are used, 
  or only low-order solutions are implemented.
  Here ``hard coded'' does not mean a compile-time adjustable parameter, 
  but that all the arrays therein are not dynamically allocated for high precision.
  (3) The software packages were not extensible, 
  \textit{i.e.}, only a single algorithm for integration was implemented in a code,
  and all parts are coupled together,
  and no interface was designed to load other codes.
  (4) Some programs lack maintenance, 
  and cannot be compiled successfully on a modern GNU/Linux operating system out-of-the-box.

  Of course, the drawbacks of the old programs are caused by historical reasons:
  lack of a high performance computer, 
  performance must take priority over maintainability.
  Lack of objected oriented program (OOP) paradigm in computational physics, 
  which came with Smalltalk in 1972, 
  and became popular with C++ (released in 1983),
  Objective-C (released in 1984) and Java (released in 1995).
  FORTRAN, which is still the most popular programming language 
  in high-performance computing, introduced OOP in Fortran 90.
  Unfortunately, FORTRAN 90 and its successors are not as popular as its predecessor, 
  FORTRAN 77.
  What makes it worse is that
  every single letter had to be recorded in punched-card format 
  and as only a limited number of characters can be used in a line,
  thus overly shortened variable names were employed used by scientists in the old days,
  which unfortunately increase the difficulty of maintaining the programs.

  We tried to construct a program friendly to both end users and developers
  in order to solve the problems above in the
  Platform of ENergetic Gases, \peng{}, 
  We wrote the programs in a modernized way.
  The codes are in C++ (following C++ 17 standard), 
  and the whole program is in OOP paradigm.
  Although we have provided the whole tool chain from PECs to transport properties,
  all the algorithms are highly decoupled,
  therefore
  experienced users (developers) can easily implement their own algorithms,
  and replace ours with little effort.
  Thus, it can serve as a platform for comparing different algorithms.  
  We also decouple the user interface and the computing part,
  so that
  \peng{} can be embedded readily into other programs.

  \section{Theory}
  \label{sec:theory}
  The macroscopic properties of dilute gases can be understood via statistical mechanics,
  in which the distribution function in the phase space is of the central role.
  For an $N$-atom system, the distribution function is of $6N+1$ dimension,
  which consists of $3N$ coordinates, $3N$ velocities, and time.
  For dilute gases, however, an approximation can be made that 
  the distribution function of the whole system can be written as
  the product of all individual gas molecules,
  \textit{i.e.}, the one-body distribution function for $i$-th specie $f_i$ in $\mu$--phase space,
  which is function of the Cartesian coordinates $\vect{r}_i$, velocities $\vect{c}_i$,
  and time $t$.

  To obtain the transport properties of a dilute binary gas mixture, 
  we begin with the Boltzmann equation, which describes the dynamics of $f_i$ \cite{McCourt2003, Chapman1991, Chapman1916}
  \begin{equation}\label{eq:boltzmann}
    \begin{cases}
      \D_0f_{0} +J_0\left(f_{0}f\right) +J_{01}\left(f_{0}f_{1}\right) =0\\
      \D_1f_{1} +J_1\left(f_{1}f\right) +J_{10}\left(f_{1}f_{0}\right) =0
    \end{cases},
  \end{equation}
  where
  $\D_i$ is referred as the differential streaming operator
  \begin{equation}
    \D_i=\pdv{t}+\vect{c}_{i}.\frac{\partial}{\boldsymbol{\partial} \vect{r}_i}+\vect{F}_{i}.\frac{\partial}{\boldsymbol{\partial} \vect{c}_i},
  \end{equation}
  in which
  $\vect{F}_i$ is external force,
  which is typically a function of coordinates.
  The $J$'s account for the encounters between gas molecules,
  where $J_0$ and $J_1$ covers the encounters between same type molecules,
  while $J_{01}$ and $J_{10}$ describe the ones between different molecules,
  \textit{e.g.}, $J_0$ and $J_{01}$ are defined as
  \begin{equation}
    \begin{split}
      J_0(f_0f)&\equiv\iiint(f_0f-f_0'f')gb\dd{b}\dd{\epsilon}\dd{\vect{c}},\\
      J_{01}(f_0f_1)&\equiv\iiint(f_0f_1-f_0'f_1')gb\dd{b}\dd{\epsilon}\dd{\vect{c}_1},
    \end{split}
  \end{equation}
  and $J_1$ and $J_{10}$ can be defined likewise.
  %We will focus on the case of $f_1$ in the following introduction, 
  %and the formulae for $f_2$ can be obtained easily by changing the suffixes.
  In the $J$'s definition,
  $g$ is the magnitude of the pre-collision relative velocity, 
  $b$ is the ``impact parameter'' of the encounter, 
  $\epsilon$ is an angle corresponding to the azimuthal orientation 
  of the scattering plane.
  A prime ($'$) indicates a function of a post-encounter 
  velocity while the one without a prime indicates a pre-encounter velocity.
  $J$ is related to intermolecular potential energy:
  microscopically,
  pre- and post-encounter velocities are the initial velocity of a direct encounter and reverse encounter,
  of which the kinetics is the same as the one of direct encounter but only of the backward direction in time,
  respectively,
  and the kinetics are of course, closely related with the intermolecular potential energy curves.

  The Boltzmann equation is a non-linear integrodifferential equation:
  \textit{e.g.}, in the equation for $f_i$, $\D_i$ accounts for the differential part while $J_i$ and $J_{ij}$ ($j=1-i$) for the integral one.
  It is non-trivial to solve such equations.
  However, effort given by Hilbert\cite{Courant1953}, Chapman\cite{Chapman1916}, and Enskog\cite{Enskog1921} have shown 
  that we can use a perturbative way to approximate the solution, 
  where $J_i$ and $J_{ij}$ can be treated as a perturbation term.
  In this manner, $f_i$, $\D_i$, $J_i+J_{ij}$ can be expended as
  \begin{equation}
    f_i = f_i^{(0)} + f_i^{(1)} + f_i^{(2)} +\ldots ,
  \end{equation}
  \begin{equation}
    \D_if_i = \D_i^{(0)} + \D_i^{(1)} + \D_i^{(2)} +\ldots ,
  \end{equation}
where $\D_i^{(r)}$  is defined as
\begin{equation}
	\D_i^{(r)}=\frac{\partial_{r-1}f_i^{(0)}}{\partial t}+\frac{\partial_{r-2}f_i^{(1)}}{\partial t}+\ldots +\frac{\partial_{0}f_i^{(r-1)}}{\partial t}
	+\left(\vect{c}_i.\frac{\partial}{\boldsymbol{\partial}\vect{r}}+\vect{F}_i.\frac\partial{\boldsymbol{\partial}\vect{c}_i}\right)f_i^{(r-1)},
\end{equation}
and
  \begin{equation}
    J_i(f_if) + J_{ij}(f_if_j)= J_i^{(0)} + J_i^{(1)} + J_i^{(2)} +\ldots ,
  \end{equation}
in which
  \begin{equation}
    \begin{split}
      J_i^{(r)}=& J_i(f_i^{(0)}f^{(r)}) + J_i(f_i^{(1)}f^{(r-1)})  +\ldots+J_i(f_i^{(r)}f^{(0)}) \\
      &+J_{ij}(f_i^{(0)}f_j^{(r)}) + J_{ij}(f_i^{(1)}f_j^{(r-1)})  +\ldots+J_{ij}(f_i^{(r)}f_j^{(0)})  .\\
    \end{split}
  \end{equation}
  These terms obey
  \begin{equation}
    \D_1^{(r)}+J_1^{(r)} =0,\quad r=0,1,2,\ldots ,
  \end{equation}
  We need also define the operator ${\partial_r}/{\partial t}$, which is given in \ref{sec:pdvr}.
  To sum up, $\D_1^{(r)}$ is only related to $f_1^{(0)}$, $f_1^{(1)}$, \ldots, and $f_1^{(r-1)}$,
  and the only term contains $f_1^{(r)}$ is related to $J_1^{(r)}$.
  Thus, Boltzmann equation can be solved step by step from an initial trial of $f_1^{(0)}$.

  In the $0$-th solution of $f_i$, \textit{i.e.}, $f_i^{(0)}$, we have $J_i^{(0)}=0$,
  which corresponds to the uniform and steady gas, the distribution function is the Maxwellian velocity distribution
  \begin{equation}\label{eq:maxwellian}
    f_i^{(0)}=n_i\left(\frac{m_i}{2\pi\kB T}\right)^{3/2}\exp{-m_i(\vect{c}_i-\vect{c})^2/2\kB T},
  \end{equation}
  where $\kB$ is the Boltzmann constant,
$T$ is the temperature, $p$ is the hydrostatic pressure,
$m_{i}$ is the mass of the $i$-th atoms,
$\vect{c}$ is the mass-weighted mean velocity of the gas,  
and 
$n_i$ is the number of specie $i$.

%  and the higher-order corrections satisfy
%  \begin{equation}
%    \begin{alignedat}{3}
%      \int f_i^{(r)} \dd\vect{c}_i &= 0, &\qquad& r=1,2,3,\ldots, \\
%      \sum_im_i\int\vect{c}_if_i^{(r)}\dd\vect{c}_i &= 0, && r=1,2,3,\ldots,\\
%      \sum_im_i\int(\vect{c}_i-\vect{c}_0)^2f_i^{(r)}\dd\vect{c}_i &= 0, && r=1,2,3,\ldots.
%    \end{alignedat}
%  \end{equation}
%  We can then solve these terms of $f_i$ step by step.

  %The $r$-th order term of $f_i$'s can be written as a product of the $0$-th order term and a function $\Phi^{(r)}$, 
  %\textit{e.g.}, 
  In the first order perturbation, $f_i^{(1)}$ can be written as
  \begin{equation}
      f_i^{(1)}=f_i^{(0)} \Phi_i^{(1)}.
  \end{equation} 
Thus,
  \begin{equation}\label{eq:I}
    \begin{split}
      \D_i^{(1)}  =&-J_i^{(1)}\\
      =&-J_i(f_i^{(0)}f^{(0)}\Phi^{(1)})
      -J_i(f_i^{(0)}\Phi_i^{(1)}f^{(0)})\\
      &-J_{ij}(f_i^{(0)}f_j^{(0)}\Phi_j^{(1)})
      -J_{ij}(f_i^{(0)}\Phi_i^{(1)}f_j^{(0)})\\
      =&-n_i^2 I_i\left(\Phi_i^{(1)}\right)-n_i n_j I_{ij}\left(\Phi_i^{(1)}+\Phi_j^{(1)}\right), 
      %\D_1^{(1)} & =-n_1^2 I_1\left(\Phi_1^{(1)}\right)-n_1 n_2 I_{12}\left(\Phi_1^{(1)}+\Phi_2^{(1)}\right), \\
      %\D_2^{(1)} & =-n_2^2 I_2\left(\Phi_2^{(1)}\right)-n_1 n_2 I_{21}\left(\Phi_1^{(1)}+\Phi_2^{(1)}\right),
    \end{split}
  \end{equation}
  %where $\D_i^{(r)}=\D f_i^{(r)}$, and
  %\begin{equation}
    %\begin{aligned}
      %\mathcal{D}_i^{(r)}=& \frac{\partial_{r-1} f_i^{(0)}}{\partial t}+\frac{\partial_{r-2} f_i^{(1)}}{\partial t}+\cdots+\frac{\partial_0 f_i^{(r-1)}}{\partial t} \\
      %&+\left(\vect{c}_i \cdot \frac{\partial}{\partial \vect{r}}+\vect{F}_i \cdot \frac{\partial}{\partial \vect{c}_i}\right) f_i^{(r-1)},
    %\end{aligned}
  %\end{equation}
  %which is the stream term of $r$-th order, and
  %\begin{equation}
    %\begin{split}
      %n_i^2I_i(F)&=\iint f_i^{(0)}f^{(0)}(F_i+F-F_i'-F')g\alpha_i\dd \vect{e}'\dd\vect{c},\\
      %n_in_jI_{ij}(K)&=\iint f_i^{(0)}f_j^{(0)}(K-K')g\alpha_{ij}\dd\vect{e}'\dd\vect{c}_j,
    %\end{split}
  %\end{equation}
  where $I_i$ and $I_{ij}$ are pre-defined integrals, and are linear functionals of their argument, whose definitions are given in \ref{sec:integrals}.
  The $\D_i^{(1)}$ can be written as 
  \begin{equation}\label{eq:ADB}
      \D_i^{(1)}=f_i^{(0)}\left\{(\mathscr{C}_i^2-\frac{5}{2})\vect{C}_i.\frac{\partial\ln T}{\vect{\partial}\vect{r}}+x_i^{-1}\vect{d}_{ij}.\vect{C}_i+2\nodiv{{\C}_i{\C}_i}:\frac{\partial\vect{c}}{\vect{\partial}\vect{r}},\right\},
  \end{equation}
where $\vect{C}_i \equiv \vect{c}_i -\vect{c}$ is the peculiar velocity,
$\C_i\equiv(m_i/2\kB T)^{1/2}\vect{C}_i$ is the dimensionless peculiar velocity,
$x_i\equiv n_i/(n_i+n_j)$ is the mole fraction of the $i$-th specie, and
\begin{equation}
	\vect{d}_{ij}=-\vect{d}_{ji}\equiv x_{i}\frac{\partial}{\boldsymbol{\partial}\vect{r}}\ln p_i-\frac{\rho_i\rho_j}{\rho p}(\vect{F}_i-\vect{F}_j)-\frac{\rho_i}{\rho p}\frac{\partial p}{\boldsymbol{\partial}\vect{r}}.
\end{equation}
where $\rho = \rho_0+\rho_1 $ is the total density, $\rho_i\equiv n_im_i$,
and $p$ is the hydrostatic pressure ($p_i$ the partial pressure).
Compare Eqs. \eqref{eq:I} and \eqref{eq:ADB}, we can give the \textit{ansatz} of $\Phi_i^{(1)}$
\begin{equation}
	\Phi_i^{(1)}=-\vect{A}_i.\frac{\partial\ln T}{\vect{\partial}\vect{r}}-\vect{D}_i.\vect{d}_{ij}-2\mat{B}_i:\frac{\partial\vect{c}}{\vect{\partial}\vect{r}},
\end{equation}
where integrals of $\vect{A}$, $\vect{D}$, and $\mat{B}$ should equal the corresponding coefficients in Eq.~\eqref{eq:ADB}, \textit{i.e.},
\begin{alignat}{2}
	f_i^{(0)}(\mathscr{C}_i^2-5/2)\vect{C}_i &= n_i^2I_i(\vect{A}_i)+n_in_jI_{ij}(\vect{A}_i+\vect{A}_j), \label{eq:A}\\
	(-1)^{i}x_i^{-1}f_i^{(0)}\vect{C}_i &= n_i^2I_i(\vect{D}_i)+n_in_jI_{ij}(\vect{D}_i+\vect{D}_j), \label{eq:D}\\
	f_i^{(0)}\nodiv{\C_i\C_i} &= n_i^2I_i(\mat{B}_i)+n_in_jI_{ij}(\mat{B}_i+\mat{B}_j).\label{eq:B}
\end{alignat}
If there is no external force acting on the system, it can be shown that
\begin{equation}
	\vect{d}_{ij}=\frac{\partial x_i}{\vect{\partial} \vect{r}} = n^{-1}\vect{\nabla}n_i.
\end{equation}

Now let us consider the transport properties by get the expectation value with the first order approximation of $f_i$, \textit{i.e.}, $f_i^{(0)}+f_i^{(1)}$.
Diffusion means that the two components of the gas have different average velocity
\begin{equation}
	\begin{split}
		\overline{\vect{C}}_0-\overline{\vect{C}}_1=&\frac{1}{n_0}\int f_0\vect{C}_0\dd{\vect{c}_0}-\frac{1}{n_1}\int f_1\vect{C}_1\dd{\vect{c}_1}\\
		=&-\frac{1}{3}\Big[\Big\{\frac{1}{n_0}\int f_0^{{0}}\vect{C}_0.\vect{D}_0\dd{\vect{c}_0}-\frac{1}{n_1}\int f_1^{{0}}\vect{C}_1.\vect{D}_1\dd{\vect{c}_1}\Big\}\vect{d}_{01} \\
		&+\Big\{\frac{1}{n_0}\int f_0^{{0}}\vect{C}_0.\vect{A}_0\dd{\vect{c}_0}-\frac{1}{n_1}\int f_1^{{0}}\vect{C}_1.\vect{A}_1\dd{\vect{c}_1}\Big\}\frac{\partial}{\vect{\partial}\vect{r}}\ln T\Big]\\
		=&\frac{1}{3}n[\{\vect{D},\vect{D}\}\vect{d}_{01}+\{\vect{D},\vect{A}\}\frac{\partial}{\vect{\partial}\vect{r}}\ln T],
	\end{split}
\end{equation}
and we define the coeefficient of diffusion as $D_{01} \equiv (n_0n_1/3n)\{\vect{D},\vect{D}\}$ and the coefficient of thermal diffusion as $D_{\textrm{T}} \equiv (n_0n_1/3n)\{\vect{D},\vect{A}\}$.
We also define the thermal-diffusion ratio $k_{\text{T}} \equiv D_{\text{T}}/D_{01}$, and thermal diffusion factor $\alpha_{01} \equiv k_{\text{T}}/x_0x_1$.

If we look into the energy flux $\vect{q}$ of the system, we can get
\begin{equation}
	\vect{q} = -\lambda \frac{\partial}{\vect{\partial}\vect{r}}T+\frac{5}{2}\kB T(n_0\overline{\vect{C}}_0+n_1\overline{\vect{C}}_1)+\kB nT k_{\text{T}}(\overline{C}_0-\overline{C}_1),
\end{equation}
where
\begin{equation}
	\lambda\equiv \frac{1}{3}\kB n^2[\{\vect{A},\vect{A}\}\{\vect{A},\vect{D}\}^2/\{\vect{D},\vect{D}\}]
\end{equation}
is the coefficient of thermal conductivity, which reflect the correspondence of gas mixture to temperature gradient.
We define $\tilde{\vect{A}}_i \equiv \vect{A}_i-k_{\text{T}}\vect{D}_i$, so
\begin{equation}
	\lambda = \frac{1}{3}\kB n^2\{\tilde{\vect{A}}, \tilde{\vect{A}}\} . %\boldsymbol{\tilde{A}}
\end{equation}
Similarly, if we consider the pressure tensor, we have
\begin{equation}
	\mat{p}^{(1)} = -\frac{4}{5}\kB n^2T\{\mat{B},\mat{B}\} \nodiv{\overline{\overline{\frac{\partial}{\vect{\partial}\vect{r}} \vect{c}}}},
\end{equation}
and the coefficient of viscosity $\mu$ is defined as 
\begin{equation}
	\label{eq:viscosity}
	\mu\equiv\frac{2}{5}\kB n^2T\{\mat{B},\mat{B}\}.
\end{equation}

Apparently, the key of getting the four coefficients is finding a robust way to get the integrals 
$\{\vect{A},\vect{A}\}$, $\{\vect{A},\vect{D}\}$, $\{\vect{D},\vect{D}\}$, and $\{\mat{B},\mat{B}\}$,
and these vectors and matrices ($\tilde{\vect{A}}$, $\vect{D}$, and $\mat{B}$) can be written as
\begin{equation}
	\begin{split}
		\tilde{\vect{A}}_0 = \sideset{}{'}\sum_{p=-\infty}^{\infty}a_p\vect{a}_0^{(p)}, &\quad
		\tilde{\vect{A}}_1 = \sideset{}{'}\sum_{p=-\infty}^{\infty}a_p\vect{a}_1^{(p)} \\
		\vect{D}_0 = \sum_{p=-\infty}^{\infty}d_p\vect{a}_0^{(p)}, &\quad
		\vect{D}_1 = \sum_{p=-\infty}^{\infty}d_p\vect{a}_1^{(p)}, \\
		{\mat{B}}_0 = \sideset{}{'}\sum_{p=-\infty}^{\infty}b_p\mat{b}_0^{(p)}, &\quad
		{\mat{B}}_1 = \sideset{}{'}\sum_{p=-\infty}^{\infty}b_p\mat{b}_1^{(p)}, 
	\end{split}
\end{equation}
where $\sideset{}{'}\sum$ means sum over all $p$ but zero, and (for $p>0$)
\begin{equation}
	\begin{split}
		\vect{a}_0^{0} = M_0^{1/2}\rho_1\C_0/\rho, &\quad \vect{a}_1^{0} = -M_1^{1/2}\rho_0\C_1/\rho, \\
		\vect{a}_0^{p} = S_{3/2}^{(p)}(\mathscr{C}_0^2)\C_0, &\quad \vect{a}_0^{-p} = \vect{0},\\
		\vect{a}_1^{p} = \vect{0}, &\quad \vect{a}_1^{-p} = S_{3/2}^{(p)}(\mathscr{C}_1^2)\C_1,\\
		\mat{b}_0^{p} = S_{5/2}^{(p-1)}(\mathscr{C}_0^2)\nodiv{\C_0\C_0}, &\quad \mat{b}_0^{-p} = \mat{0},\\
		\mat{b}_1^{p} = \mat{0}, &\quad \mat{b}_1^{-p} = S_{5/2}^{(p-1)}(\mathscr{C}_1^2)\nodiv{\C_1\C_1},\\
	\end{split}
\end{equation}
%\begin{equation}
%	\tilde{\vect{A}}_i=\C_i\tilde{A}_i(\mathscr{C}_i),\quad \vect{D}_i=\C_iD_i(\mathscr{C}_i),\quad\mat{B}=\nodiv{{\C}_i{\C}_i}B_i(\mathscr{C}_i),
%\end{equation}
%and $A_i$, $D_i$, and $B_i$ are all function of $\mathscr{C}_i$.
%We can expand these functions as linear combinations of 
where $S_m^{(n)}$ are Sonine polynomials (or in more common literature, associated Laguerre polynomials).
Sonine polynomials can be written as
\begin{equation}
	S_m^{(n)}(x) = \sum_{p=0}^{n}\frac{(-x)^p}{p!}\binom{m+n}{m-p},
\end{equation}
and they have orthogonality, \textit{i.e.}
\begin{equation}
	\int_0^{\infty}\mathrm{e}^{-x}S_m^{(p)}(x)S_m^{(q)}(x)x^m\dd{x}=
	\begin{cases}
		0& (p\neq q)\\
		\Gamma(m+p+1)/p! & (p=q)
	\end{cases}.
\end{equation}
Using Sonine polynomials as bases, 
the solution to the integrals $\{\vect{A},\vect{A}\}$, $\{\vect{A},\vect{D}\}$, $\{\vect{D},\vect{D}\}$, and $\{\mat{B},\mat{B}\}$ 
turns into linear algebra problems.
Here, we use only the case of $\mat{B}$ as an example.
We define
\begin{equation}
	\label{eq:beta}
	\{\mat{B},\mat{b}^{(p)}\} = \beta_p,
\end{equation}
and recalling Eq.~\eqref{eq:B}
%\begin{equation}
	\begin{gather}
			\beta_1 = \frac{5n_0}{2n^2}= n^{-1}\frac{5x_0}{2}, \quad
		\beta_{-1} = \frac{5n_1}{2n^2}= n^{-1}\frac{5x_1}{2},\\
		\beta_{q} =0\quad(q\neq \pm 1).\label{eq:betaeq0}
	\end{gather}
%\end{equation}
Thus,
\begin{equation}
	\begin{split}
		\{\mat{B},\mat{B}\} =& \sum_pb_p\{\mat{B},\mat{b}^{(p)}\}\\
		=&b_1\beta_{1}+b_{-1}\beta_{-1}.
	\end{split}
\end{equation} 
Here note that we use $S_{5/2}^{(p)}$ to expand $\mat{B}$ leading to $\beta_q\ 0 (q\neq \pm1)$,
because the expression of $\{\mat{B},\mat{b}^{(q)}\}$
has the structure of 
\begin{equation*}
	\int_{0}^{\infty}\exp{-\mathscr{C}^2}(\mathscr{C}^2)^{5/2}S_{5/2}^{(q-1)}(\mathscr{C}^2)\dd{(\mathscr{C}^2)},
\end{equation*}
which consists with the orthogonal property of Sonine polynomials.
$b_p$'s are then the key to get $\{\mat{B},\mat{B}\}$.
We further expand $\mat{B}$ in $\{\mat{B},\mat{b}^{q}\}$ to turn Eq.~\eqref{eq:beta} as
\begin{equation}
	\sideset{}{'}{\sum}_{p=-\infty}^{\infty} b_pb_{pq}= \beta_q,
\end{equation}
where
\begin{equation}
	b_{pq} = \{\mat{b}^{(p)},\mat{b}^{(q)}\}.
\end{equation}
This can be turned into a matrix multiplication considering we can only use limited $p$ and $q$.
As defined in Eq.~\eqref{eq:viscosity}, the resulting value of viscosity will not be dependent on the total density $n$ but will be on $x_i$,
because the $n^2$ in Eq.~\eqref{eq:viscosity} and the $n$'s from $b_i$ and $\beta_p$ cancel.
We will thus move the $n$'s to the final expression of these properties.
(Actually, in the four properties studied,
the diffusion and thermal diffusion are related to the density by a inverse proportional relation,
while viscosity and thermal conductivity are independent on $n$.)
Clearly, the other integrals can be solved in this fashion.
The matrix elements $a_{pq}$ and $b_{pq}$ can then be given using the $\Omega$ integrals, see below.

%if we expend the expression of $\D_i^{(r)}$.

  %Mathematically, we can 

%  Here, we will skip the intermediate process,
%  and give the Chapman--Enskog solution \cite{Chapman1916, Enskog1917, Chapman1991} 1
%  directly in terms of Sonine polynomials. 
%  The reader may check Refs.~\cite{Tompson2009, Tompson2010} for the detailed derivation. 
%  Note that some mistakes and missing equations in the reference 
%  have been corrected and supplemented in this work.

  Formulae for diffusion $D_{01}$ and thermal diffusion $D_{\text{T}}$ are
  \begin{equation}
    \begin{split}
      \left[D_{01}\right]_{n}=&\left(\frac{\kB T}{p}\right)\frac{1}{2}x_{0}x_{1}\left(\frac{2\kB T}{m}\right)^{1/2}d_{0}^{(n)},\\	
      \left[D_{\mathrm{T}}\right]_{n}=&-\left(\frac{\kB T}{p}\right)\frac{5}{4}x_{0}x_{1}\\
      &\times\left(\frac{2\kB T}{m}\right)^{1/2}\left(x_{0}M_{0}^{-1/2}d_{1}^{(n)}+x_{1}M_{1}^{-1/2}d_{-1}^{(n)}\right).	\\
    \end{split}
  \end{equation}
  Readers may find the equation here have additional ${\kB T}/{p} = n^{-1}$ 
  factor compared with the ones in Ref.~\cite{Tompson2009}, 
  while we remove the $n^{-1}$ from the definition of $\delta_0$ (see below).
  Diffusion is inversely proportional to the total density or pressure,
  and in practice people
  report these diffusion properties under $p=\SI{1}{atm}$, see, \textit{e.g.}, Ref.~\cite{Kestin1984}.
  In \peng, we follow this convention and print the diffusion coefficients under $\SI{1}{atm}$.

  In equations above,
  subscript $n$ ($[\cdots]_n$) means ``the $n$-th order approximation of $\cdots$'', 
  and superscription $(n)$ means element of the $n$-th order vectors (see below).
We also have the $n$-th order vectors
  \begin{equation}
    \begin{alignedat}{8}
      \vect{d}^{(n)}=	(   & d_{-n}, & \cdots, & d_{-1}, & d_{0},      & d_{1}, & \cdots, & d_{n}&)^{\T}, \\
      \vect{\delta}^{(n)}=	( & 0,      & \cdots, & 0,      & \delta_{0}, & 0,     & \cdots, & 0&)^{\T},
    \end{alignedat}
  \end{equation}
  where the superscript $\T$ means transpose,
  \begin{equation}
    \delta_{0}=\frac{3}{2}\left(\frac{2 \kB T}{m_{0}}\right)^{1 / 2}
  \end{equation}
  and
  \begin{equation}
    \vect{d}^{(n)} = \left(\vect{D}^{(n)}\right)^{-1}\vect{\delta}^{(n)}.
  \end{equation}
  The $\vect{D}^{(n)}$ matrix is
  \begin{equation}
    %\begin{split}
    \vect{D}^{(n)}%&
    =\left(\begin{array}{lll>{\columncolor{gray!20}}llll}
      a_{-n-n} & \cdots & a_{-n-1} & {a_{-n0}} & a_{-n1} & \cdots & a_{-nn} \\
      \vdots   & \ddots & \vdots   & {\vdots}  & \vdots  & \iddots & \vdots  \\
      a_{-1-n} & \cdots & a_{-1-1} & {a_{-10}} & a_{-11} & \cdots & a_{-1n} \\ 
      \rowcolor{gray!20} a_{0-n}  & \cdots & a_{0-1}  & a_{00}    & a_{01}  & \cdots & a_{0n}  \\ 
      a_{1-n}  & \cdots & a_{1-1}  & {a_{10}}  & a_{11}  & \cdots & a_{1n}  \\
      \vdots   & \iddots & \vdots   & {\vdots}  & \vdots  & \ddots & \vdots  \\
      a_{n-n}  & \cdots & a_{n-1}  & {a_{n0}}  & a_{n1}  & \cdots & a_{nn}
    \end{array}\right), %\\
    %&=\raisebox{-0.45\height}{\includegraphics{windrose.pdf}},
    %\end{split}
    \label{eq:Dmatrix}
  \end{equation}
  whose elements are
  \begin{equation}
    \begin{split}
      a_{pq}=a_{qp}     & =x_{0}^{2}H_{pq}(0000)+x_{0}x_{1}H_{pq}(0001), \\
      a_{p-q}=a_{-qp}   & =x_{0}x_{1}H_{pq}(0101),                       \\
      a_{-pq}=a_{q-p}   & =x_{1}x_{0}H_{pq}(1010),                       \\
      a_{-p-q}=a_{-q-p} & =x_{1}^{2}H_{pq}(1111)+x_{1}x_{0}H_{pq}(1110), \\
      a_{p0}=a_{0p}     & =x_{0}x_{1}M_{0}^{1/2}H_{p0}(0001),            \\
      a_{-p0}=a_{0-p}   & =-x_{1}x_{0}M_{1}^{1/2}H_{p0}(1110),           \\
      a_{00}            & =8x_{0}x_{1}M_{0}M_{1}\Omega_{01}^{(1,1)},
    \end{split}
  \end{equation}
  where $H_{pq}(ab\alpha\beta)=\left[S_{3/2}^{(p)}(\mathscr{C}_{a}^{2})\C_{a},S_{3/2}^{(q)}(\mathscr{C}_{b}^{2})\C_{b}\right]_{\alpha\beta}$,
  for any positive $p$ and $q$. Note that the last three lines were missed in Ref.~\cite{Tompson2009}. 
  The index $a$ and $b$ are for mass
  and $\alpha$ and $\beta$ designate the collision integrals $\Omega_{\alpha\beta}^{(\ell,s)}(T)$. 
  When $a=b=\alpha=\beta$, 
  \begin{equation}
    H_{pq}(ab\alpha\beta)=8\sum_{\ell=2}^{(\min[p,q]+1)}\sum_{s=\ell}^{(p+q+2-\ell)}A_{pqs\ell}'''\Omega_{\alpha\beta}^{(\ell,s)},
  \end{equation}
  where
  \begin{equation}
    \begin{split}
      A_{pqs\ell}''' =& \left(\frac{1}{2}\right)^{(p+q+1)}\sum_{i=(\ell-1)}^{\min[p,q,s,(p+q+1-s)]}\frac{8^{i}(p+q-2i)!}{(p-i)!(q-i)!}\\
      & \times\frac{1+(-1)^{\ell}}{\ell!(i+1-\ell)!}\frac{(-1)^{(s+i)}}{(s-i)!(p+q+1-i-s)!}\\
      & \times\frac{(s+1)!}{(2s+2)!}\frac{(2(p+q+2-i))!}{(p+q+2-i)!}\frac{2^{2s}}{4^{(p+q+1)}}\\
      & \times\left[(i+1-\ell)(p+q+1-i-s)-\ell(s-i)\right].
    \end{split}
  \end{equation}
  When $a\neq b,\alpha\ne\beta,a=\alpha\,\text{and}\,b=\beta$,
  \begin{equation}
    H_{pq}(ab\alpha\beta)=8\sum_{\ell=1}^{(\min[p,q]+1)}\sum_{s=\ell}^{(p+q+2-\ell)}A_{pqs\ell}''\Omega_{\alpha\beta}^{(\ell,s)},
  \end{equation}
  where
  \begin{equation}
    \begin{split}
      A_{pqs\ell}''  =&M_{b}^{(p+1/2)}M_{a}^{(q+1/2)}\\
      & \times\sum_{i=(\ell-1)}^{\min[p,q,s,(p+q+1-s)]}\frac{8^{i}(p+q-2i)!}{(p-i)!(q-i)!}\\
      & \times\frac{(-1)^{\ell}}{\ell!(i+1-\ell)!}\frac{(-1)^{(s+i)}}{(s-i)!(p+q+1-i-s)!}\\
      & \times\frac{(s+1)!}{(2s+2)!}\frac{(2(p+q+2-i))!}{(p+q+2-i)!}\frac{2^{2s}}{4^{(p+q+1)}}\\
      & \times\{(i+1-\ell)(p+q+1-i-s)-\ell(s-i)\}.
    \end{split}
  \end{equation}
  When $a=b$ while $\alpha\ne\beta$,
  \begin{equation}
    H_{pq}(ab\alpha\beta)=8\sum_{\ell=1}^{(\min[p,q]+1)}\sum_{s=\ell}^{(p+q+2-\ell)}A_{pqs\ell}'\Omega_{\alpha\beta}^{(\ell,s)},
  \end{equation}
  where
  \begin{equation}
    \begin{split}
      A_{pqs\ell}'  =&\sum_{i=(\ell-1)}^{\min[p,q,s,(p+q+1-s)]}\frac{8^{i}(p+q-2i)!}{(p-i)!(q-i)!}          \\
      & \times\frac{1}{\ell!(i+1-\ell)!}\frac{(-1)^{(s+i)}}{(s-i)!(p+q+1-i-s)!}                \\
      & \times\frac{(s+1)!}{(2s+2)!}\frac{(2(p+q+2-i))!}{(p+q+2-i)!}\frac{2^{2s}}{4^{(p+q+1)}} \\
      & \times\sum_{w=0}^{(\min[p,q,(p+q+1-s)]-i)}F^{(i+1-\ell)}\frac{G^{w}}{w!}               \\
      & \times\frac{(p+q+2-i-s-w)_{w}}{(2(p+q+2-i)-2w+1)_{w}}(p+1-i-w)_{w}                     \\
      & \times\frac{(p+q+3-i-w)_{w}}{(2(p+q+2-i)-w+1)_{w}}(q+1-i-w)_{w}                        \\
      & \times2^{(2w-1)}\frac{M_{a}^{i}M_{1-a}^{i}M_{1-a}^{(p+q-2i-w)}}{(p+q+1-2i-w)_{w}}      \\
      & \times\Big[2M_{a}F^{-1}(i+1-\ell)(p+q+1-i-s-w)                                         \\
      & -2M_{1-a}\ell(s-i)\Big],
    \end{split}
  \end{equation}
  in which $F=\left(M_{0}^{2}+M_{1}^{2}\right)/2M_{0}M_{1}$ and $G=\left(M_{a}-M_{1-a}\right)/M_{1-a}$.
  In formulae above, $n!=\Gamma(n+1)$ is the factorial, 
  and $\left(z\right)_n = \Gamma(z+n)/\Gamma(z)$ is the Pochhammer symbol. 
  Note that we employ a zero-based index so that subscript $1-a$ represents the mass not appearing
  explicitly in $a$ or $b$.

  The thermal conductivity $\lambda$ of the binary mixture is given as
  \begin{equation}
    \begin{split}
      \left[\lambda\right]_{n}  =&-\frac{5}{4}\kB \left(\frac{2\kB T}{m}\right)^{1/2}      \\
      & \times\left(x_{0}M_{0}^{-1/2}a_{0}^{(n)}+x_{1}M_{1}^{-1/2}a_{-1}^{(n)}\right).
    \end{split} 
  \end{equation}
  The rule of super- and sub-scripts is the same as that for diffusion, 
  and $\vect{a}^{(n)}$ and $\vect{\alpha}^{(n)}$  here are
  \begin{equation}
    \begin{alignedat}{9}
      \vect{a}^{(n)}=	(   &  a_{-n}, &  \cdots, &  a_{-1},      &  a_{1},      &  \cdots, &  a_{n}&)^{\mathsf{T}}, \\
      \vect{\alpha}^{(n)}=	( &  0,      &  \cdots, &  \alpha_{-1}, &  \alpha_{1}, &  \cdots, &  0&)^{\mathsf{T}},
    \end{alignedat}
  \end{equation}
  where
  \begin{equation}
    \alpha_{-1}=-\frac{15}{4} x_1\left(\frac{2 \kB T}{m_{1}}\right)^{1 / 2}, \alpha_{1}=-\frac{15}{4} x_0 \left(\frac{2 \kB T}{m_{0}}\right)^{1 / 2},
  \end{equation}
  and
  \begin{equation}
    \vect{a}^{(n)}=\left(\mat{A}^{(n)}\right)^{-1}\vect{\alpha}^{(n)}.
  \end{equation}
  Here, $\mat{A}^{(n)}$ is the $(0,0)$ minor matrix of $\mat{D}^{(n)}$, 
  \textit{i.e.}, remove the colored row and column in Eq.~\ref{eq:Dmatrix}.

  The viscosity $\eta$ is similarly given by
  \begin{equation}
    [\eta]_{n}=\kB T\left(x_{1} b_{1}^{(n)}+x_{2} b_{-1}^{(n)}\right),
  \end{equation}
  where $N$ is the number density of the gas mixture, 
  \begin{equation}
    \begin{array}{rlrrrrrrr}
      \vect{b}^{(n)}=&	(&b_{-n},& \cdots,&b_{-1},&b_{1},&\cdots,&b_{n})^{\mathsf{T}},\\
      \vect{\beta}^{(n)}=&	(&0,& \cdots,&\beta_{-1},&\beta_{1},&\cdots,&0)^{\mathsf{T}},	
    \end{array}
  \end{equation}
  while
  \begin{equation}
    \vect{b}^{(n)}=\left(\vect{B}^{(n)}\right)^{-1}\vect{\beta}^{(n)}.
  \end{equation}
  The expressions for elements in $\vect{\beta}$ are 
  (note that the redundant negative signs in Ref.~\cite{Tompson2010} have been removed) 
  \begin{equation}
    \beta_{1}=\frac{5}{2} x_0, \quad \beta_{-1}=\frac{5}{2} x_1.
  \end{equation}

  The shape of the matrix $\mat{B}^{(n)}$ looks the same as  $\mat{A}^{(n)}$, namely,
  \begin{equation}
    %\begin{split}
    \vect{B}^{(n)}%&
    =\left(\begin{array}{llllll}
      b_{-n-n} & \cdots & b_{-n-1} & b_{-n1} & \cdots & b_{-nn} \\
      \vdots   & \ddots & \vdots   & \vdots  & \iddots & \vdots  \\
      b_{-1-n} & \cdots & b_{-1-1} & b_{-11} & \cdots & b_{-1n} \\
      b_{1-n}  & \cdots & b_{1-1}  & b_{11}  & \cdots & b_{1n}  \\
      \vdots   & \iddots & \vdots   & \vdots  & \ddots & \vdots  \\
      b_{n-n}  & \cdots & b_{n-1}  & b_{n1}  & \cdots & b_{nn}
    \end{array}\right), %\\
    %&=\raisebox{-0.45\height}{\includegraphics{windrose.pdf}},
    %\end{split}
  \end{equation}
  with elements ($p>0$ and $q>0$)
  \begin{equation}
    \begin{split}
      b_{pq}=a_{qp}     & =x_{0}^{2}L_{pq}(0000)+x_{0}x_{1}L_{pq}(0001), \\
      b_{p-q}=a_{-qp}   & =x_{0}x_{1}L_{pq}(0101),                       \\
      b_{-pq}=a_{q-p}   & =x_{1}x_{0}L_{pq}(1010),                       \\
      b_{-p-q}=a_{-q-p} & =x_{1}^{2}L_{pq}(1111)+x_{1}x_{0}L_{pq}(1110). 
    \end{split}
  \end{equation}
  The $L_{pq}(ab\alpha\beta)$ is defined as
  \begin{equation}
    L_{pq}(ab\alpha\beta)  = \left[S_{5 / 2}^{(p-1)}\left(\mathscr{C}_{a}^{2}\right) \nodiv{\C_{a} \C_{a}}, S_{5 / 2}^{(q-1)}\left(\mathscr{C}_{b}^{2}\right) \nodiv{\C_{b} \C_{b}}\right]_{\alpha\beta}.
  \end{equation}

  When $a=b=\alpha=\beta$,
  \begin{equation}
    L_{pq}(ab\alpha\beta)=\frac{16}{3} \sum_{\ell=2}^{(\min [p, q]+2)} \sum_{s=\ell}^{(p+q+4-\ell)} B_{p q s \ell}''' \Omega_{\alpha\beta}^{(\ell,s)},
  \end{equation}
  where
  \begin{equation}
    \begin{split}
      B_{p q s \ell}''' & =\left(\frac{1}{2}\right)^{(p+q+2)} \frac{2^{2 s}}{4^{(p+q+2)}}                                                                           \\
      & \times \frac{(s+1) !}{(2 s+2) !} \frac{\left[1+(-1)^{\ell}\right]}{\ell !} \sum^{\min [p, q, s,(p+q+2-s)]}_{i=(\ell-2)}  \\
      & \times \frac{(-1)^{s+i}}{(p+q+2-i-s) !} \frac{(p+q-2 i) !}{(p-i) !(q-i) !(s-i) !}                                                     \\
      & \times \frac{[2(p+q+3-i)] !}{(p+q+3-i) !} \frac{8^{i}}{(i+2-\ell) !}\\
      & \times \Big\{(i+1-\ell)(i+2-\ell)\\                                          
      & \times\big[(p+q+1-i-s)(p+q+2-i-s)\\
      & - \tfrac{1}{2}(s-i)(s-i-1)\big]\\
      & +\tfrac{3}{2}(\ell-1) \ell(s-i)(s-i-1) \\ 
      & -2 \ell(i+2-\ell)(s-i)(p+q+2-i-s)\Big\}.
    \end{split}
  \end{equation}
  When $a\neq b,\alpha\ne\beta,a=\alpha\,\text{and}\,b=\beta$,
  \begin{equation}
    L_{pq}(ab\alpha\beta)=\frac{16}{3} \sum_{\ell=1}^{(\min [p, q]+2)} \sum_{s=\ell}^{(p+q+4-\ell)} B_{p q s \ell}'' \Omega_{
      \alpha\beta}^{(\ell,s)},
  \end{equation}
  where
  \begin{equation}
    \begin{split}
      B_{p q s \ell}'' & = M_{b}^{(p+1)} M_{a}^{(q+1)}\left(1-\delta_{s,(p+q+3)}\right)                                                            \\
      & \times \sum_{i=(\ell-2)}^{\min [p, q, s,(p+q+2-s)]} \frac{2^{2 s}}{4^{(p+q+2)}} \frac{8^{i}(p+q-2 i) !}{(p-i) !(q-i) !}   \\
      & \times \frac{(-1)^{(s+i)}\left(1-\delta_{i,(-1)}\right)}{(s-i) !(p+q+2-i-s) !}                                            \\
      & \times \frac{(s+1) !}{(2 s+2) !} \frac{[2(p+q+3-i)] !}{(p+q+3-i) !}  \frac{(-1)^{\ell}}{\ell !(i+2-\ell) !}               \\
      & \times \Big\{(i+1-\ell)(i+2-\ell)\\                                                                                
      & \times \big[(p+q+1-i-s)(p+q+2-i-s)\\
      & -\tfrac{1}{2}(s-i)(s-i-1)\big]\\ 
      & +\tfrac{3}{2}(\ell-1) \ell(s-i)(s-i-1)\\
      & -2 \ell(i+2-\ell)(s-i)(p+q+2-i-s)\Big\} .
    \end{split}
  \end{equation}
  When $a=b$ while $\alpha\ne\beta$,
  \begin{equation}
    L_{pq}(ab\alpha\beta)=\frac{16}{3} \sum_{\ell=1}^{(\min [p, q]+2)} \sum_{s=\ell}^{(p+q+4-\ell)} B_{p q s \ell}' \Omega_{\alpha\beta}^{(\ell,s)},
  \end{equation}
  where
  \begin{equation}
    \begin{split}
      &B_{p q s \ell}'=\left(1-\delta_{s,(p+q+3)}\right)\\
      &\times \sum_{i=(\ell-2)}^{\min [p, q, s,(p+q+2-s)]} \frac{2^{2 s}}{4^{(p+q+2)}} \frac{8^{i}(p+q-2 i) !}{(p-i) !(q-i) !}\\
      &\times \frac{(-1)^{(s+i)}\left(1-\delta_{i,(-1)}\right)}{(s-i) !(p+q+2-i-s) !} \frac{(s+1) !}{(2 s+2) !} \\
      & \times \frac{(2(p+q+3-i)) !}{(p+q+3-i) !} \\ 
      &\times \sum_{w=0}^{(\min [p, q,(p+q+2-s)]-i)} \frac{(p+1-i-w)_{w}(q+1-i-w)_{w}}{w !(p+q+1-2 i-w)_{w}}\\
      &\times \frac{(p+q+3-i-s-w)_{w}}{(2(p+q+3-i)-2 w+1)_{w}} 2^{(2 w-2)} G^{w}\\
      &\times \frac{(p+q+4-i-w)_{w}}{(2(p+q+3-i)-w+1)_{w}} \\
      &\times M_{a}^{i} M_{1-a}^{i} M_{1-a}^{(p+q-2 i-w)} \frac{F^{(i+2-\ell)} M_{a}^{2}}{\ell !(i+2-\ell) !} \\
      &\times 4\Big\{\frac{3}{2} \frac{M_{1-a}^{2}}{M_{a}^{2}} \ell(\ell-1)(s-i)(s-i-1)\\
      &-\frac{2}{F} \frac{M_{1-a}}{M_{a}} \ell(i+2-\ell)(s-i)(p+q+2-i-s-w)\\
      &+\frac{1}{F^{2}}(i+1-\ell)(i+2-\ell)\\
      &\times \big[(p+q+1-i-s-w)(p+q+2-i-s-w)\\
      &-\frac{1}{2} \frac{M_{1-a}^{2}}{M_{a}^{2}}(s-i)(s-i-1)\big]\Big\}.
    \end{split}
  \end{equation}
  with $F=\left(M_{0}^{2}+M_{1}^{2}\right)/2M_{0}M_{1}$ and $G=\left(M_{a}-M_{1-a}\right)/M_{1-a}$.

  In the formulae above, 
  the central components are the collision integrals $\Omega_{\alpha\beta}^{(\ell,s)}$, 
  which are determined by classical collisions between atom pairs.
  The numerical computation of $\Omega_{\alpha\beta}^{(\ell,s)}$ 
  is often performed in reduced units,
  where lengths are given as multiples of the collision radius $\sigma$ 
  and energies are multiples of the well depth $\varepsilon$ of the PEC,
  \textit{e.g.}, the reduced PEC is $V^{\star}(r^{\star})=\varepsilon V(\sigma r^{\star})$,
  in which the star (${}^{\star}$) indicates a dimensionless reduced quantity.
  Clearly, 
  we can have all the quantities involved in classic collision in 
  the ``star'' format by using the reduced units introduced above.

  To compute $\Omega_{\alpha\beta}^{(\ell,r)}$ numerically, three integration must be carried out.
  The scattering angle $\chi$ is 
  a function of initial kinetic energy of the collision $E^{\star}$ and the collision parameter $b^{\star}$, and is given by
  \begin{equation}
    \begin{split}
      \chi(E^{\star},b^{\star}) & =\pi-2b^{\star}\int_{r_{\mathrm{m}}}^{\infty}\frac{\dd{r^{\star}}}{r^{{\star}2}F(r^{\star})},\\
      &\text{with}\quad b^{\star}\in[0,+\infty),E^{\star}\in[0,+\infty).
    \end{split}    
    \label{eq:chi}
  \end{equation}
  with
  \begin{equation}
    F(r^{\star})=\left[1-\frac{V^{\star}(r^{\star})}{E^{\star}}-\frac{b^{{\star}2}}{r^{{\star}2}}\right]^{1/2}.
  \end{equation}
  %we want to have everything in $r$. In short, we express $b$ in $r_{\mathrm{m}}$,
  %\begin{equation}
  %b=r_{\mathrm{m}}\sqrt{\frac{E-V(r_{\mathrm{m}})}{E}}
  %\end{equation}
  %we have $\chi=\chi(E,b)=\chi(E,r_{\mathrm{m}}).$
  %We further have $E=V(r_{\mathrm{E}})$. We should point out we assume
  %the potential energy function monotonically decrease in $(0, \sigma]$,
  %which is a fair assumption. Thus,
  %\[
  %\chi=\chi(E,b)=\chi(E,r_{\mathrm{m}})=\chi(r_{\mathrm{E}},r_{\mathrm{m}}),\quad r_{\mathrm{E}}\in(0,\sigma],r_{\mathrm{m}}\in[r_{\mathrm{E}},\infty)
  %\]
  %Similarly, we transform the formulae for $Q^{(\ell)}(E)$,
  The cross section $Q^{(\ell){\star}}$ is a function of the initial kinetic energy $E^{\star}$, 
  \begin{equation}
    Q^{(\ell){\star}}(E^{\star})  =\cfrac{2}{1-\frac{1+(-1)^{\ell}}{2(1+\ell)}}\int_{0}^{\infty}\left(1-\cos^{\ell}\chi(E^{\star},b^{\star})\right)b^{\star}\dd{b^{\star}}.
    \label{eq:q}
  \end{equation}
  The collision integral $\Omega^{{\star}(\ell,s)}$ is function of the reduced temperature $T^{\star}$
  \begin{equation}
    \Omega^{\star(\ell,s)}(T^{\star})=\frac{1}{(s+1)!}\int_{0}^{\infty}e^{-x}x^{s+1}Q^{{\star}(\ell)}(xT^{\star})\dd{x},
    \label{eq:omega}
  \end{equation}
  and $\Omega^{(\ell,s)}(T)$ can be obtained using the expression
  \begin{equation}
    \Omega^{(\ell,s)}(T)=\frac{\frac{1}{2}(s+1)!\left[ 1-\frac{1+(-1)^{\ell}}{2(1+\ell)}\right]}{\sqrt{2\pi\mu/kT}}\Omega^{\star(\ell,s)}(T^{\star}).
  \end{equation}
  Different numerical integration schemes can be implemented for these three integrals.
  Also, different schemes can be combined freely. 

  \section{Program Implementation and Developers' Guide}
  \label{sec:impl}

  \subsection{Structure of \peng{}}
\begin{figure}[h]
    \centering
    \includegraphics[width=\linewidth]{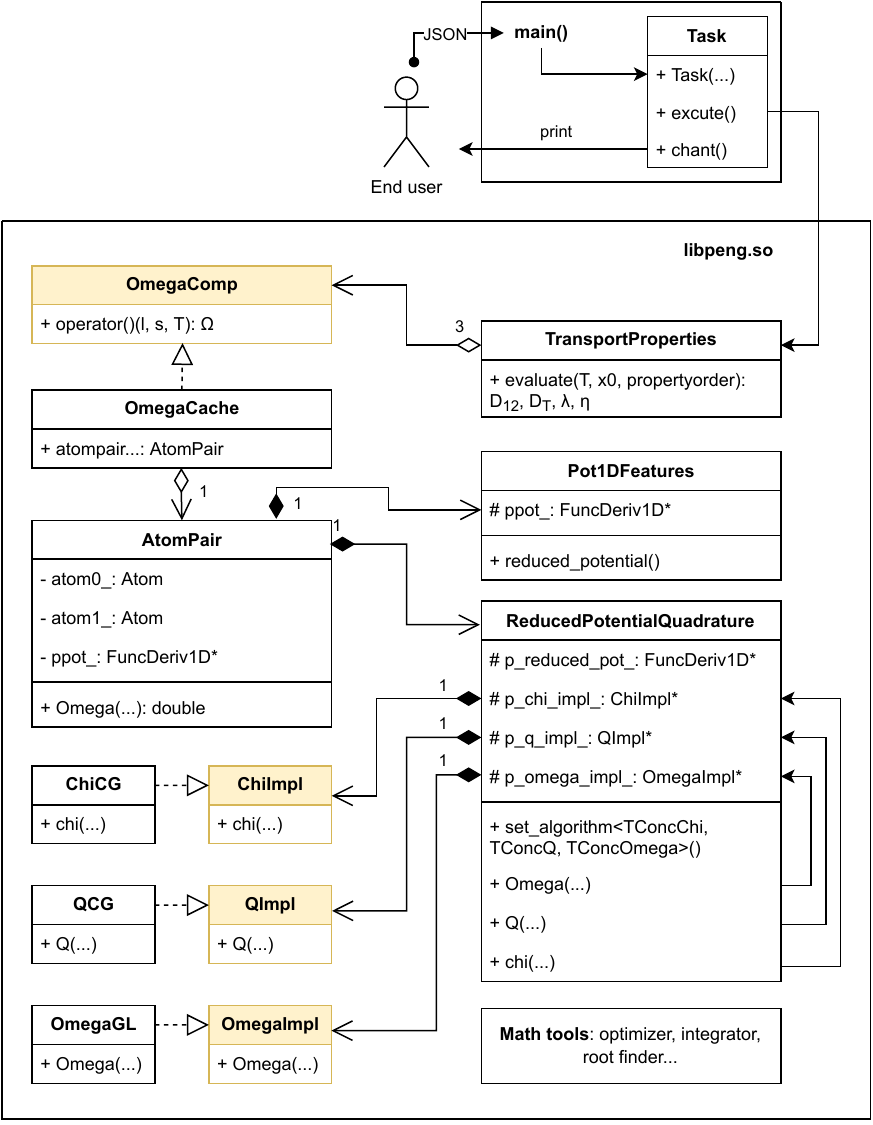}
    \caption{The Unified Modeling Language (UML) diagram of \peng{}.  
    Only the main components of the program are plotted.
    The extensible ones are colored in light orange.\label{fig:uml}}
  \end{figure}

  The main structure of the program is shown in Fig.~\ref{fig:uml}.
  The program is designed in Object-Orient Program (OOP) paradigm.
  Before proceeding, some explanation about OOP should be given.
  Although it is often taught in books about OOP \cite{Lippman1999, Lippman2013, Metcalf2018} 
  that objects (classes) are abstractions of objects (or things) in real life,
  it is a common practice \cite{Eastman2017, Tribello2014} to present an algorithm as an object.
  Alternatively, readers may think such objects abstraction of computers of specific usage.  
  For those readers who are familiar with procedural programming,
  \textit{e.g.}, Fortran 77 users,
  OOP is an elegant way to have common variables,
  and multiplex the same code for different problems 
  within the same program with minimal effort. 

  All codes in the core part of \peng{} are in the namespace \texttt{peng}.
  In this work, 
  atom pair (class \texttt{AtomPair}) is the basic component in our mental model.
  For binary mixture of atoms $0$ and $1$,
  three kinds of pair should be considered: atom pair $00$, $01$, and $11$.
  \texttt{AtomPair} class hold the basic information of the atoms in pair,
  which is recorded in \texttt{Atom} class,
  and the interatomic PEC, as a \texttt{FuncDeriv1D} object.

  The \texttt{FuncDeriv1D} is an abstract for one-dimension functions $f(x)$ together with their derivatives.
  The most important two  methods of this class are \texttt{value()}
  and \texttt{derivative()}.
  In the other part of the program, 
  we always keep the calling to these two methods with the same $x$ together, 
  if both value and derivative are needed.
  Thus, 
  a cache mechanism may be used to store the intermediate variables 
  to save computational time.
  Note that \texttt{FuncDeriv1D::derivative()}
  is not a pure virtual function, 
  the program will fallback to a default numerical differential program
  if no analytical one is provided.

  The original potential function will be passed to 
  \texttt{Pot1DFeatures} to get $\sigma$ and $\varepsilon$
  numerically, and a reduced potential $V^{\star}$ will be returned by function
  \texttt{Pot1DFeatures::reduced\_potential()}.
  $V^{\star}$ will then be passed to 
  class \texttt{ReducedPotentialQuadrature}
  for to get $\Omega^{(\ell,s)\star}(T^{\star})$,
  which is given by its \texttt{Omega()} method,
  which calls its \texttt{Q()} and
  \texttt{Chi()} methods automatically.

  The three key methods of \texttt{Reduced\-Potential\-Qua\-dra\-ture}, 
  say, \texttt{Omega()},
  \texttt{Q()}, and
  \texttt{chi()}, 
  are sockets of the three classes,
  \texttt{OmegaImpl}, \texttt{QImpl}, and \texttt{ChiImpl}. 
  Implementations of these three pure virtual classes can be loaded at compiling time
  with the template method \texttt{set\_algorithm<TChi, TQ, TOmega>()} 
  of \texttt{Reduced\-Potential\-Qua\-dra\-ture}.
  For developers, these concrete classes are where one can extend \peng{}.
  Classes \texttt{OmegaGL}, \texttt{QCG}, and \texttt{ChiCG}
  are provided as examples of integration algorithms.

  Once we have the program for $\Omega^{(\ell,s)}(T)$ ready,
  We may also obtain the transport properties, 
  which are coded in class \texttt{TransportProperties}.
  Although we have our own $\Omega^{(\ell,s)}(T)$,
  we still leave an interface, class \texttt{OmegaComp},
  and a concrete class \texttt{OmegaCache},
  which uses \texttt{AtomPair} only when new combinations 
  of $\ell$, $s$, and $T$ are passed in, or returns
  the cached value of $\Omega^{(\ell,s)}(T)$ from a local database otherwise,
  is implemented.
  This is because we do not want to limit the user to our implementations
  of collision integrals.
  Besides, $\Omega^{(\ell,s)}(T)$ are computed on-the-fly, 
  to avoid unnecessary high-order computations,
  which are more difficult to compute than low-order ones.

  Although \peng{} is designed as a library,
  a user interface (UI) is provided for end users.
  The UI is designed to be separated from the library,
  therefore, the codes of UI provide a good example for developers of
  how to use \peng{} in their own projects. 

  \subsection{Algorithms implemented in \peng{}}

  To give the end users a usable software,
  and to test the framework,
  we have supplied some concrete classes for integration.
  Following Ref.~\cite{Taylor1979}, 
  $\chi$ is rewritten as a function of $r_{\mathrm{m}}$,
  \begin{equation}
    \chi(E^{\star},r^{\star}_{\mathrm{m}})=\chi\Big(E^{\star},b^{\star}=r^{\star}_{\mathrm{m}}\sqrt{\frac{E^{\star}-V^{\star}(r^{\star}_{\mathrm{m}})}{E^{\star}}}\Big),
  \end{equation}
  and the integration in $[r_{\mathrm{m}},\infty)$ is mapped to $(0,1)$ 
  by $r\to \frac{r_{\mathrm{m}}}{r}$, 
  and the Chebyshev--Gauss (CG) quadrature rule \cite{Abramowitz1972} is used.
  Note that CG rule is for $(-1,1)$.
  To use it under this circumstance, 
  we assume the integrand of the $chi$ quadrature is an even function,
  and we only take the positive half.
  The CG quadrature rule is based on Chebyshev nodes,
  which are nested. 
  In our implementation, $3^n$ ($n = 1,2,\ldots$) nodes are used,
  and the integration algorithm is written in class \texttt{CGQuadrature}.

  The cross-section $Q^{(\ell)\star}$ is 
  \begin{equation}
    \begin{split}
      Q^{(\ell)\star}(E^{\star})  =&\cfrac{2}{1-\frac{1+(-1)^{\ell}}{2(1+\ell)}}\int_{0}^{\infty}\left(1-\cos^{\ell}\chi\right)b^{\star}\dd{b^{\star}}                                                                            \\
       =&\cfrac{1}{1-\frac{1+(-1)^{\ell}}{2(1+\ell)}}\int_{0}^{\infty}\left(1-\cos^{\ell}\chi\right)\dd(b^{\star 2})                                                                                   \\
       =&\cfrac{1}{1-\frac{1+(-1)^{\ell}}{2(1+\ell)}}\left[\int_{r^{\star}_{\mathrm{E}}}^{r^{\star}_{\mathrm{O'}}}\!\!+\!\!\int_{r^{\star}_{\mathrm{O}}}^{\infty}\right]\left(1-\cos^{\ell}\chi\right) \\
      & \times \dd\left(r^{\star 2}_{\mathrm{m}}{\frac{E^{\star}-V^{\star}(r^{\star}_{\mathrm{m}})}{E^{\star}}}\right).
    \end{split}
  \end{equation}
  Due to orbiting \cite{Taylor1979, Hirschfelder1954, OHara1971,Barker1964},
  the integration must be divided into two parts when the energy is less
  than $E^{\star}_{\mathrm{C}}$, corresponding to $r^{\star}_{\mathrm{C}}$, 
  which is the minimum of the function 
  \begin{equation}
    y(r^{\star})=V^{\star}(r^{\star})+\tfrac{1}{2}r^{\star}\derivative{V^{\star}(r^{\star})}{r^{\star}}.
  \end{equation}
  $r'_{\mathrm{E}}$ is the smallest for a specific $E^{\star}$,
  corresponding to $b^{\star} = 0$.
  $r^{\star}_{\mathrm{O}}$ and $r^{\star}_{\mathrm{O}'}$ are roots of
  \begin{equation}
    V^{\star}(r^{\star})+\frac{b^{{\star}2} E^{\star}}{r^{{\star}2}} -E^{\star} =0.
  \end{equation}
  When $E^{\star}$ is larger than $E^{\star}_{\mathrm{C}}$,
  $r^{\star}_{\mathrm{O}}=r^{\star}_{\mathrm{O'}}=r^{\star}_{\mathrm{C}}$.
  The two integration may be carried out numerically using the CG quadrature rule,
  and have already been written in class \texttt{QCG}.
  Note that when orbiting happens,
  i.e., $|\chi|$ is large (in the present implement, $|\chi|>20\pi$),
  it is very difficult to obtain an accurate scattering angle numerically.
  Following Ref.~\cite{Barker1964},
  $\cos^l(\chi)$ is replaced by the average value
  \begin{equation}
    \cos^{\ell}(\chi)=\begin{cases}
      0\quad &(\ell\text{ is odd}) \\
      \frac{(\ell-1)!!}{\ell!!}=\frac{\Gamma(\frac{\ell}{2}+\frac{1}{2})}{\Gamma(\frac{\ell}{2}+1)\sqrt{\pi}}\quad &(\ell\text{ is even})
    \end{cases}.
  \end{equation}

  To compute $\Omega^{(\ell,s)\star}$, 
  generalized Gauss--Laguerre (GL) quadrature rule \cite{Abramowitz1972} is used,
  because the formula for $\Omega^{(\ell,s)\star}$ is in the format of the GL rule,
  of which the grids and weight 
  are given by an automatic quadrature rule generator \cite{Burkardt2010}.
  Note that in generalized GL rule, different quadrature grids are used when $s$ changes.
  The computed $\Omega^{(\ell,s)\star}(T^{\star})$,
  implemented in \texttt{OmegaGL},
  is then called by class \texttt{AtomPair} 
  to obtain $\Omega^{(\ell,s)}(T)$.

  We notice that with our algorithms,
  the computational wall time is slightly longer than those of Refs.~\cite{Barker1964, OHara1971, Taylor1979}.
  This is mainly because we did not take the interpolation approximation of $Q$.
  In the reference implementations, $Q$ used by $\Omega$ is obtained from interpolation.
  However, we find that this made the final results unstable if 
  transport properties at higher temperature are required,
  in which case the interpolation pivots may be changed.
  Of course, this problem can be solved by using a nested interpolation grids
  with spline, for example.
  However, we do have more powerful computers now, 
  so direct computation of $Q$ is affordable.
  Should the user require such an interpolation, 
  one could embed our \texttt{QCG} class in their implementation of \texttt{QImpl},
  or write their own.

  \subsection*{}

  All other details are written as inline comments in \peng{}'s source code,
  and the HTML format documentation can be generated using Doxygen \cite{Doxygen}.
  Developers can follow these documents should they wish to add new algorithms. 
  Beside the syntax check provided by compilers, 
  we use Valgrind \cite{Nethercote2007} to check 
  that \peng{} does not suffer from problems from memory leakage or uninitialized variables.

  \section{End Users' Guide}
  \label{sec:enduser}

  \subsection{Potential Energy Curves}
  Firstly, ``end users'' should still be able to write simple C code,
  because \peng{} needs to call user-provided PECs.
  Due to name mangling, 
  the lite way to employ user-provided PECs is to use a C interface,
  because it is \textit{de facto} standard on GNU/Linux.

  At least one function providing the value of PEC should be written, 
  of which the signature is (in C)
\begin{verbatim}
double value(double r);
\end{verbatim}
For a real system, \texttt{r} should be in Angstrom 
and the return value in Kelvin.

It is often the case where the analytical derivative of the PEC 
can be provided. In such a case, also write the function
\begin{verbatim}
double derivative(double r);
\end{verbatim}
Do not provide an incorrect derivative 
when knowledge of the derivative is not available. 
If the program did not find the derivative function, 
it will compute the numerical derivative instead
(see class \texttt{FuncDeriv1D} in the previous section). 

The PECs for the three atom pairs should be in three separated files.
For example, 
if we wish to tell \peng{} that we have a Lennard--Jones Potential,
we write the following codes (in file \texttt{lj.c}),
\begin{lstlisting}[language=C]
/**
 * File: lj.c
 * Lennard-Jones Potential (12-6)
 */

#include <math.h>

/* Some local cache */
static double rm6 = 0.0;
static double oldr = -1.0;

void update_r(double r) {
  if (oldr != r) {
    rm6 = pow(r, -6);
    oldr = r;
  }
}

/* V(r)=4(r^-12-r^-6) */
double value(double r) {
  update_r(r);
  return 4 * (rm6 * rm6 - rm6);
}

/* V'(r)=4(-12*r^-13+6r^-7) */
double derivative(double r) {
  update_r(r);
  return 4 * (-12 * rm6 * rm6 / r + 6 * rm6 / r);
}
\end{lstlisting}
and compile it as a dynamic library
\begin{lstlisting}[language=sh]
gcc -O3 -Wall -fPIC -shared -o lj.so lj.c 
\end{lstlisting}

\subsection{JSON input file}
Other information, 
alongside the position of the prepared PECs, 
are written in input files.
The input file are in JSON format,
which is a popular human-readable format to record 
multiple types of data.
\textit{E.g.}, the JSON file for \ce{He} and \ce{Xe} takes the form
\begin{lstlisting}[]
{
  "atoms": [
    {
      "name": "He",
      "mass": 4.00260325413
    },
    {
      "name": "Xe",
      "mass": 131.9041550856
    }
  ],
  "potentials":[
    {
      "name": "HeHe",
      "path": "./hehe.so"
    },
    {
      "name": "HeXe",
      "path": "./hexe.so"
    },
    {
      "name": "XeXe",
      "path": "./xexe.so"
    } 
  ],
  "accuracy": 1.0e-2,
  "temperatures": [
    50.0,
    100.0,
    150.0,
    200.0,
    250.0,
    300.0
  ],
  "temperatures.bak": [
    50.0
  ],
  "propertyorder": 3,
  "molefractions0": [
    0.25,
    0.50,
    0.75
  ],
  "molefractions0.bak": [
    0.25
  ]
}
\end{lstlisting}
The sample JSON file is self-explained, 
and we list the keys below. 
If unmentioned keys are written in the input file, 
\textit{e.g.}, \texttt{temperatures.bak} in the sample file, 
they will not affect the computation, 
thus, they can be used as comments.
\begin{enumerate}
  \item \texttt{atoms}: Array of the two types of atoms, name is not really used, and mass in atomic mass unit [amu, $\frac{1}{12} m(\ce{^{12}C})$].
  \item \texttt{potentials}: Array of the PECs, in the order of interactions between atoms[0]-atoms[0], atoms[0]-atoms[1], and atoms[1]-atoms[1].
  \item \texttt{accuracy}: Maximum of allowed integration relative error.
  \item \texttt{temperatures}: Array of the temperatures at which the properties are computed.
  \item \texttt{molefractions0}: Array of the mole fractions of atoms[0], and the mole fraction of atoms[1] is computed accordingly.
  \item \texttt{propertyorder}: The maximum order when compute the transport properties.	
\end{enumerate}

Once the JSON input file 
and the dynamical libraries of the PECs are prepared,
The user can run \peng{} with 
\begin{lstlisting}[language=sh]
/path/to/peng.exe < input.json 
\end{lstlisting}
Pre-formatted tables,
with heads being elements and their mole fractions,
as well as the approximation order of the computation,
following with values of the computed results,
will be printed to \texttt{stdout} with names and units of the results.
Intermediate information, will be printed to \texttt{stderr}.
The user needs a modern terminal support UTF-8 encoding to show the tables
because \peng{} uses some Unicode characters in the unit literals.
Users may find examples in the \texttt{examples} directory of the source codes.

\section{Case study: binary mixture of \ce{He} and \ce{Xe}}

With \peng, we can easily compute the transport properties from the first principle,
which can be compared with experimental data.
If the potential energy functions are based on high-level electronic structure theories
and reasonable models,
then the results should be reliable,
and the theory can be used to confirm the experiments.
\textit{E.g.}, for the binary mixture of \ce{He} and \ce{Xe},
we use three potential energy curves for the pairs \ce{He\bond{...}He}\cite{Przybytek2010}, \ce{Xe\bond{...}Xe}\cite{Hellmann2017}, and \ce{He\bond{...}Xe}\cite{Liu2022}.
These three state-of-the-art potential energy curves are very accurate,
in which high-order coupled cluster theory are used with large basis sets and relativistic corrections has been made.
For intermolecular interactions, the methodology used can provide spectroscopically accurate results.
Thus, thermophysical prediction based on these potential energy curves is a very important reference for understanding theoretical data.

\begin{figure}[h]
	\centering
	\includegraphics{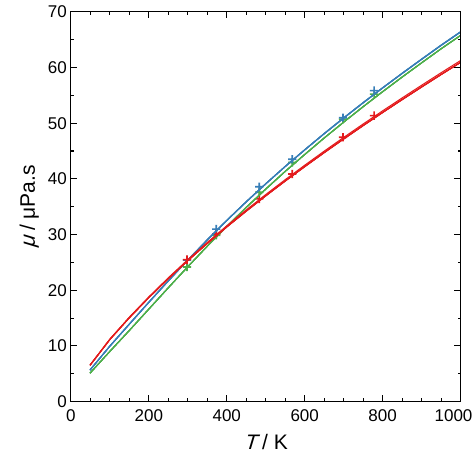}
	\caption{Viscosity of \ce{He} and \ce{Xe} mixture.
		Red: $x_{\ce{He}} = 0.8687$, blue: $x_{\ce{He}} = 0.5316$, and green: $x_{\ce{He}} = 0.2677$.
		The lines are computed results of this work and the pluses are measured data from Ref.~\cite{Kestin1978}.
		Viscosity curves with different approximation are plotted with the same color, so lines of the same color overlap. 
		\label{fig:kestin}}
\end{figure}

\begin{figure}[h]
	\centering
	\includegraphics{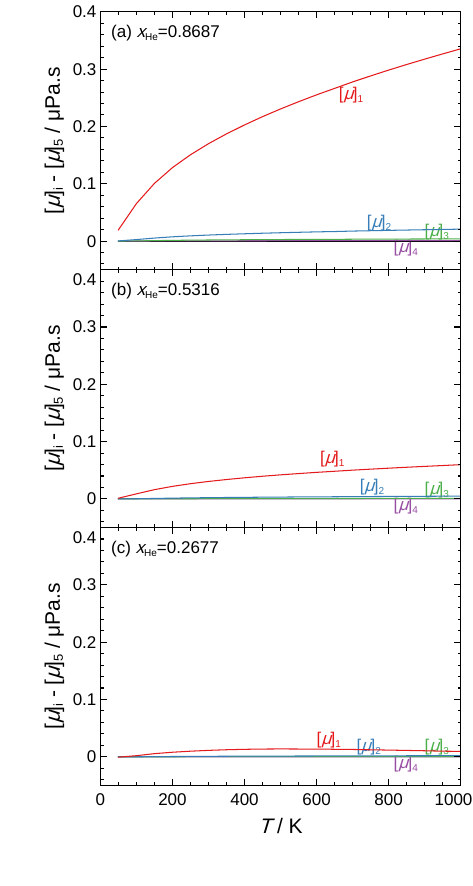}
	\caption{Convergence of viscosity of \ce{He} and \ce{Xe} mixture.
		Panel a: $x_{\ce{He}} = 0.8687$, panel b: $x_{\ce{He}} = 0.5316$, and panel c $x_{\ce{He}} = 0.2677$.
		For different mole fraction, the convergence are different.
		\label{fig:kestin_conv}}
\end{figure}

Because the masses of \ce{He} and \ce{Xe} differ with each other greatly, 
higher-order approximation of the Chapman--Enskog solution is needed.\cite{Storvick1966}
Viscosity of \ce{He} and \ce{Xe} mixture is interesting.
Experiment\cite{Kestin1978} showed that the viscosity-temperature curves for mixtures of different mole fraction cross,
and theoretical computation can confirm such cross at $\sim\SI{300}{K}$ (Fig.~\ref{fig:kestin}).
Here, we compute the viscosity to the fifth order.
The curves converge quickly in general.
Noticeably, viscosity for the mixture with mole fraction of \ce{He} being $0.8687$ converges slower than the other two do, 
$[\mu]_4$ is an accurate approximation (Fig.~\ref{fig:kestin_conv}a), 
while for $x_{\ce{He}} = 0.2677$, even $[\mu]_1$ is acceptable (Fig.~\ref{fig:kestin_conv}c).

\begin{figure}[h]
	\centering
	\includegraphics{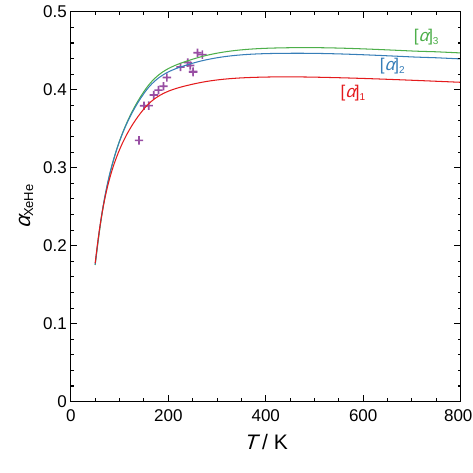}
	\caption{Thermal diffusion factor of \ce{He} and \ce{Xe} equimolar mixture.
		Red: first order; blue: second order; and green: third order approximation of $\alpha$.
		The purple pluses are experimental data from \cite{Hurly1991}.
		\label{fig:hurly}}
\end{figure}

We can also compute the thermal diffusion factor $\alpha_{01}$ based on coefficients for diffusion and thermal diffusion, 
and compare them with available experimental data.
In Fig.~\ref{fig:hurly}, the calculated $\alpha$ is is plotted.  
Interestingly, the $\alpha_{\ce{XeHe}}(T)$ function is not monotonic.
The calculated curve agrees with the experimental data\cite{Hurly1991}.
However, such agreement is not very satisfactory.
Considering the deviation between different experimental works (see Fig.~2 in Ref.~\cite{Hurly1991}),
computational data may be more reliable in this case.
We computed the diffusion coefficients to the third order of Sonine polynomial expansion of distribution function,
and find that the curve converge uniformly.
Also note that to get the curve, high accurate $\Omega$ integrals are needed.
In this case, we cannot get a smooth curve unless a threshold of $1\times10^{-5}$ for integrals are used.

\section{Summary and Outlook}

We presented \peng{},
an extensible C++ library that compute the collision integrals between 
atom pairs,
and transport properties 
(diffusion, thermal diffusion, thermal conductivity, and viscosity)
based on the Chapman--Enskog solution of Boltzmann equation.

When designing the program, 
extensibility is kept in mind. 
The main three steps of computing classical collision integrals,
angle of deflection $\chi$, cross-section $Q$, 
and collision integral $\Omega$
can be implemented as concrete class of virtual classes
\texttt{ChiImpl}, \texttt{QImpl}, and \texttt{OmegaImpl}.
Although we provide usable implementations based on Chebyshev--Gauss
and Gauss--Laguerre quadrature rules,
however, we will implement more algorithms, 
and contributions from the community are welcomed.
Users can also only use the \texttt{Trans\-portPro\-per\-ties} class
to use with their own integration code with minimal effort.
\peng{} gives the user the ability of computing 
thermophysical properties,
and the method developers a platform to test their algorithms.

Of course, the present version of \peng{} can only compute
classical collision integrals, 
and can only deal with monatomic gas binary mixtures,
although these functions are rare in computer programs 
publicly available.
In the future, more functions, 
\textit{e.g.}, computing quantum collision integrals \cite{Hirschfelder1954, Taylor1986},
dealing with polyatomic gases \cite{McCourt1990}, 
and more flexible quadrature rules 
(including doing the integration on multi-dimension sparse grids)
can be added in newer version.

\section*{Acknowledgment}
Hui Li thanks the late Professor Robert J. Le Roy (University of Waterloo, Canada)
for long-term inspiring discussions.
This work has been supported by the 2020-JCJQ project (GFJQ2126-007)
and the National Natural Science Foundation of China (Grant No.: 22073035),
and The Program for JLU Computational Interdiscipline 
Innovative Platform of Jilin University, China.

\section*{Conflicts of interest}

The authors declare no competing financial interest.

\section*{Author contributions}
\begin{itemize}
  \item Y.Z.: Investigation; Methodology; Software; Visualization; Writing - original draft.
  \item Y.L.: Software; 
  \item H.L.: Supervision; Methodology; Software; Writing - review \& editing.
  \item F.R.W.M.: Supervision; Methodology; Writing - review \& editing.
\end{itemize}

\appendix
\section{Build Guide}
\label{sec:build}

\peng{} is easy to install.
For end users, 
with a GNU/Linux computer connected to the Internet,
simply type
\begin{lstlisting}[language=sh]
make -j
\end{lstlisting}
in the root of the source code,
a Bash script \texttt{external/prerequest.sh}
will run to download the packages needed by \peng{},
and build them automatically before compiling \peng{}.
For cases in which there is no Internet access,
one needs the following packages
\begin{enumerate}
  \item fmt (\url{https://fmt.dev/latest/index.html}),
  \item JSON for modern C++ (\url{https://json.nlohmann.me/}),
  \item NLOpt (\url{https://nlopt.readthedocs.io/en/latest/}),
  \item Eigen (\url{https://eigen.tuxfamily.org/}).
\end{enumerate}
Please find the detailed compiling flags in the shell script.

The code can be compiled
with \texttt{GCC 12.1.1}, \texttt{Clang 14.0.6}, and \texttt{Intel ICC 2021.6.0}
on an openSUSE x86\_64 computer.
It also works on Android (with Termux) and Microsoft Windows (with Windows Subsystem for Linux, WSL).

\section{Mathematical preparations}
Readers, especially those who do not familiar with the statistical theory for gases,
may find this paper hard to understand because the mathematics.
In this section, we will give the definitions of symbols, and some of their properties.
We will, however, not provide the proof because of the length of this work.

\subsection{Scalars, vectors, and tensors}\label{sec:linealgebra}

In this work, following the tradition of \cite{Chapman1991}, we use italic font ($a$) to represent a scalar, 
and bold italic font ($\vect{a}$) to denote a vector.
\textit{E.g.}, velocity $\vect{c}=(u,v,w)$ is a vector, whose $x$, $y$, and $z$ components are $u$, $v$, and $w$.
So a function $f$ of $\vect{c}$ would be of $u$, $v$, and $w$.
If we want to get the gradient of $f$, it would be
\begin{equation}
	\frac{\partial}{\vect{\partial}\vect{c}}f = \left(\pdv{f}{u},\pdv{f}{v},\pdv{f}{w}\right).
\end{equation}

The scalar (inner) product of two vectors are written as $\vect{a}.\vect{b}$, 
which gives a scalar equals the sum of the products of each pair of the compoments of $\vect{a}$ and $\vect{b}$.
The divergence of a function $f$ can be, therefore, denoted by
\begin{equation}
	\frac{\partial}{\vect{\partial}\vect{c}}.f = \pdv{f}{u}+\pdv{f}{v}+\pdv{f}{w}.
\end{equation}

If the dot does not present, $\vect{a}\vect{b}$ is a second order tensor (matrix),
which is denoted by bold {Sans Serif} font ($\mat{w}$)
\begin{equation}
	\mat{w} = \vect{a}\vect{b}=
	\begin{pmatrix}
		a_xb_x & a_xb_y & a_xb_z \\		a_yb_x & a_yb_y & a_yb_z \\		a_zb_x & a_zb_y & a_zb_z \\
	\end{pmatrix}=
\begin{pmatrix}
	w_{xx} & w_{xy} & w_{xz} \\		w_{yx} & w_{yy} & w_{yz} \\		w_{zx} & w_{zy} & w_{zz} \\
\end{pmatrix}
\end{equation}
The divergence of a tensor  is the sum of its diagonal elements, it is easy to turn a tensor as non-divergence,
and is denoted by $\nodiv{\mat{w}}$
\begin{equation}
	\nodiv{\mat{w}} = \mat{w} - \frac{1}{3}(w_{xx}+w_{yy}+w_{zz}) \mat{U}
\end{equation}
for a tensor in $xyz$ space,
where $\mat{U}$ is the unitary matrix.
If the rows and columns of a tensor $\mat{w}$ is interchanged, we have the transposed one $\overline{\mat{w}}$.
The symmetricalized $\mat{w}$ is then 
\begin{equation}
	\overline{\overline{\mat{w}}}=\frac{1}{2}(\mat{w}+\overline{\mat{w}}).
\end{equation}
This symbol, can be combined with the non-divergence symbol, as $\nodiv{\overline{\overline{\mat{w}}}}$.

Simple product of tensors is still a tensor, whose elements are 
\begin{equation}
	(\mat{w}.\mat{w}')_{\alpha\beta} = \sum_{\gamma}w_{\alpha\gamma}w_{\gamma\beta}',
\end{equation}
while the double, or scalar product is 
\begin{equation}
	\mat{w}:\mat{w}'= \sum_{\alpha\beta}w_{\alpha\beta}w_{\beta\alpha}'. 
\end{equation}

\subsection{${\partial_r }/{\partial t}$ in $\D^{(r)}$}
\label{sec:pdvr}
The $r$-th order partial derivative of a function $F$ is defined using the chain rule
\begin{equation}
	\frac{\partial_r F}{\partial t} = \sum_{\lambda\in \{n, \vect{c}, T\}}\pdv{F}{\lambda}\frac{\partial_r \lambda}{\partial t}.
\end{equation}
We define ${\partial_r \lambda}/{\partial t}$ as, respectively,
\begin{alignat}{2}
		\frac{\partial_r n}{\partial t} =& 
		\begin{cases}
			\displaystyle-\frac{\partial}{\vect{\partial} \vect{r}}.(n\vect{c})&(r=0)\\
			0&(r>0)
		\end{cases},\\
			\frac{\partial_r \vect{c}}{\partial t} =& 
	\begin{cases}
		\displaystyle-\left(\vect{c}.\frac{\partial}{\vect{\partial}\vect{r}}\right)\vect{c}+\vect{F}-\frac{1}{\rho}\frac{\partial}{\vect{\partial}\vect{r}}.\mat{p}^{(0)}&(r=0)\\
		\displaystyle-\frac{1}{\rho}\frac{\partial}{\vect{\partial}\vect{r}}.\mat{p}^{(r)}&(r>0)
	\end{cases},\\
			\frac{\partial_r T}{\partial t} =& 
\begin{cases}
	\displaystyle-\vect{c}.\frac{\partial T}{\vect{\partial}\vect{r}}-\frac{2}{3\kB n}\left(\mat{p}^{(0)}:\frac{\partial }{\vect{\partial}\vect{r}}\vect{c}+\frac{\partial }{\vect{\partial}\vect{r}}.\vect{q}^{(0)}\right)&(r=0)\\
	\displaystyle-\frac{2}{3\kB n}\left(\mat{p}^{(r)}:\frac{\partial }{\vect{\partial}\vect{r}}\vect{c}+\frac{\partial }{\vect{\partial}\vect{r}}.\vect{q}^{(r)}\right)&(r>0)
\end{cases}.
\end{alignat}

\subsection{Integrals related to $f^{(0)}$}\label{sec:integrals}
Again, we write the Maxwellian velocity distribution function, 
\textit{i.e.}, the distribution function for a uniform and steady gas (Eq.~\eqref{eq:maxwellian})
\begin{equation*}
	f_i^{(0)}=n_i\left(\frac{m_i}{2\pi\kB T}\right)^{3/2}\exp{-m_i(\vect{C}_i)^2/2\kB T}.
\end{equation*}

Let $F$, $G$, $H$, $K$, \ldots{} be functions of velocity (we will label them with subscripts to specify the variable), 
and we define three kinds of integrals.
$I$ is defined as
\begin{equation}
	\begin{split}
		n_0^2I_0(F) &\equiv \iiint f_0^{(0)} f^{(0)}(F_0+F-F_0'-F')gb\dd{b}\dd{\epsilon}\dd{\vect{c}},\\
		n_0n_1I_{01}(K) &\equiv \iiint f_0^{(0)} f_1^{(0)}(K-K')gb\dd{b}\dd{\epsilon}\dd{\vect{c}},
	\end{split}
\end{equation}
Clearly, the $I$ integrals are linear with respect to their variables,
and note that $I$ is only partly integrated, thus $I_0$ and $I_{01}$ are actually functions of $\vect{c}_{0}$,
and likewise $I_1$ and $I_{10}$ are of $\vect{c}_{1}$.

We can also construct complete integrals.
The bracket integral $[F,G]$ are defined as 
\begin{equation}
	[F,G]_0=[G,F]_0\equiv \int G_0 I_0(F)\dd{\vect{c}_0}.
\end{equation}
For $F$ and $H$ are arbitrary function of $\vect{c}_0$ and $G$ and $K$ are of $\vect{c}_1$,
we define
\begin{equation}
	\begin{split}
	[F_0+G_1, H_0+K_1]_{01}\equiv& \int F_0 I_{12}(H_0+K_1)\dd{\vect{c}_0}+\int G_1 I_{21}(H_0+K_1)\dd{\vect{c}_1}\\
	=&[H_0+K_1, F_0+G_1]_{01}.
	\end{split}
\end{equation}
For binary mixture, we also use the following compound integral ($F$ and $G$ are both functions of $\vect{c}_0$ and $\vect{c}_1$)
\begin{equation}
	\begin{split}
		n^2\{F,G\} =& n^2\{G,F\} \\
		 \equiv &n_0^2[F,G]_0 +n_0n_1[F_0+F_1, G_0+G_1]_{01}+n_1^2[F,G]_1.
	\end{split}
\end{equation}
It can be proved that $[F,G]$ and $\{F,G\}$ are linear with respect to the two variables $F$ and $G$.

%% References


\bibliographystyle{elsarticle-num}
\bibliography{000ref.bib}






% \end{thebibliography}

\end{document}